# Refractive Index–Correlated Pseudocoloring for Adaptive Color Fusion in Holotomographic Cytology

Minseok Lee,[1,6] Tal Lifshitz,[1,4] Young Ki Lee,[2,3] Geon Kim,[1,4] Seog Yun Park,[2,3] Hayoung Lee,[2,3] Juyeon Park[1,4], Eun Kyung Lee[2,3,*] YongKeun Park[1,4,5,*]

[1]Department of Physics, Korea Advanced Institute of Science and Technology (KAIST), South Korea;

[2]Department of Internal Medicine, National Cancer Center, South Korea;

[3]Department of Pathology, National Cancer Center, South Korea;

[4]KAIST Institute for Health Science and Technology, KAIST, Daejeon, Republic of Korea

[5]Tomocube Inc., Daejeon, Republic of Korea

[6]Current affiliation: Department of Electrical Engineering & Computer Science, Massachusetts Institute of Technology, Cambridge, MA 02139, USA

Corresponding authors: E.K. Lee (eklee@ncc.re.kr) and Y.K. Park (yk.park@kaist.ac.kr)

Received: XX October 2025 | Revised: XX October 2025 | Accepted: XX October 2025

Funding: This work was supported by National Research Foundation of Korea grant funded by the Korea government (MSIT) (RS-2024-00442348, 2022M3H4A1A02074314), Korea Institute for Advancement of Technology (KIAT) through the International Cooperative R&D program (P0028463), and the Korean Fund for Regenerative Medicine (KFRM) grant funded by the Korea government (the Ministry of Science and ICT and the Ministry of Health & Welfare) (21A0101L1-12).

## Abstract

Conventional bright-field (BF) cytology of thyroid fine-needle aspiration biopsy (FNAB) suffers from staining variability and limited subcellular contrast. Here, we present a refractive index–correlated pseudocoloring (RICP) framework that integrates quantitative refractive index (RI) maps obtained by holotomography (HT) with color BF images to enhance diagnostic interpretability. The imaging platform combines a digital micromirror device (DMD)–based HT system with an RGB LED illumination module, enabling simultaneous acquisition of RI tomograms and BF images from PAP-stained thyroid samples. The RICP algorithm adaptively embeds RI-derived structural information into the least-occupied hue channel, preserving color fidelity while enhancing nuclear and cytoplasmic contrast. Applied to benign and malignant thyroid clusters, RICP revealed diagnostically relevant features such as nucleoli, lipid droplets, and nuclear irregularities, and hue–saturation analysis quantitatively differentiated cytological categories. This perceptually grounded, label-free framework bridges conventional color cytology and quantitative optical imaging for improved diagnostic precision.

# 1. Introduction

Cytological examination remains a cornerstone of clinical diagnostics, providing rapid and minimally invasive assessment of cellular morphology in fine-needle aspiration (FNA) specimens (1). Conventional bright-field (BF) microscopy, combined with chemical staining, enables visualization of nuclear and cytoplasmic features that guide diagnostic interpretation. However, this technique is inherently limited—contrast depends on staining uniformity and optical absorption, while subtle intracellular features such as nucleoli, lipid droplets, or chromatin texture often remain obscured (2). In addition, inter-observer variability and staining artifacts can compromise diagnostic reproducibility.

Quantitative phase imaging, particularly holotomography (HT), provides a label-free method for visualizing cellular architecture by measuring the refractive index (RI), which reflects local biomolecular density and composition (3,4). This RI information complements traditional color-based cytology by providing objective and quantitative optical contrast. In particular, three-dimensional (3D) reconstruction of RI distributions using HT has enabled detailed visualization and analysis of cellular morphology (5), including lipid droplets (LDs) (6–8), liquid–liquid phase separation (9–11), cellular membranes (12), immune cell morphology (13–16), volumetric 3D tissues (17,18), and organoids (19,20). Nevertheless, HT images lack the intuitive color cues familiar to cytologists, which limits their direct interpretability in routine clinical workflows.

Color fusion across multiple imaging modalities has long been explored as a means to combine complementary structural and functional information into a single interpretable image. In medical imaging, conventional fusion approaches typically operate in perceptual color spaces such as hue–saturation–value (HSV) or intensity-hue-saturation (HIS), where structural modalities (e.g., MRI or CT) are injected into the intensity or value channel while preserving the hue from functional modalities such as PET or SPECT (21–23). Similar strategies have been employed in microscopy to overlay bright-field and fluorescence images using fixed channel assignments or alpha-blending rules (24–26), as well as in hyperspectral and pan-sharpening applications where a high-resolution intensity component replaces the luminance channel of RGB images (27,28). These approaches enhance visibility but often rely on *static* channel mappings that do not account for the local color occupancy or perceptual balance of the base image, leading to color distortion and suboptimal information integration.

Recent efforts have sought to improve perceptual consistency by optimizing color mapping based on human color appearance models (29), by preserving hue statistics during multimodal fusion (30), and deep learning–based multimodal fusion of imaging (31). However, existing methods still treat the color-space allocation as fixed, rather than adaptively identifying the least-occupied or complementary color region for information embedding. To our knowledge, no previous study has leveraged *statistical hue analysis* of bright-field images to determine an optimal complementary channel for fusing physically distinct image modalities—such as RI tomograms and bright-field images—into a unified, diagnostically meaningful pseudocolor representation.

To address this gap, we developed a refractive index–correlated pseudocoloring (RICP) framework that adaptively embeds RI-derived structural contrast into the least-occupied hue channel of a BF image. This adaptive color fusion preserves the native chromaticity of stained cytological specimens while introducing orthogonal RI-based contrast that enhances diagnostically relevant subcellular features, including nuclear membranes, nucleoli, and cytoplasmic granularity such as lipid droplets. The proposed method is demonstrated using human thyroid FNAB samples imaged by correlative HT, which simultaneously acquires BF and 3D RI maps. Each pixel's three BF color channels and one RI value are transformed into a new three-channel representation optimized for perceptual consistency and quantitative integrity.

Figure 1 illustrates the conceptual comparison between conventional cytological imaging and the proposed RICP framework. In standard BF cytology (top), FNA smears are stained and imaged to produce color images with limited contrast and potential artifacts. In the proposed workflow (bottom), the same samples are imaged via holotomography to acquire both BF and RI data, which are then fused through RICP to generate a complementary pseudocolor image that combines the interpretability of BF microscopy with the quantitative specificity of RI contrast. This framework establishes a bridge between conventional color cytology and label-free quantitative imaging, offering a perceptually grounded and physically interpretable approach for multimodal cytological visualization and diagnosis.

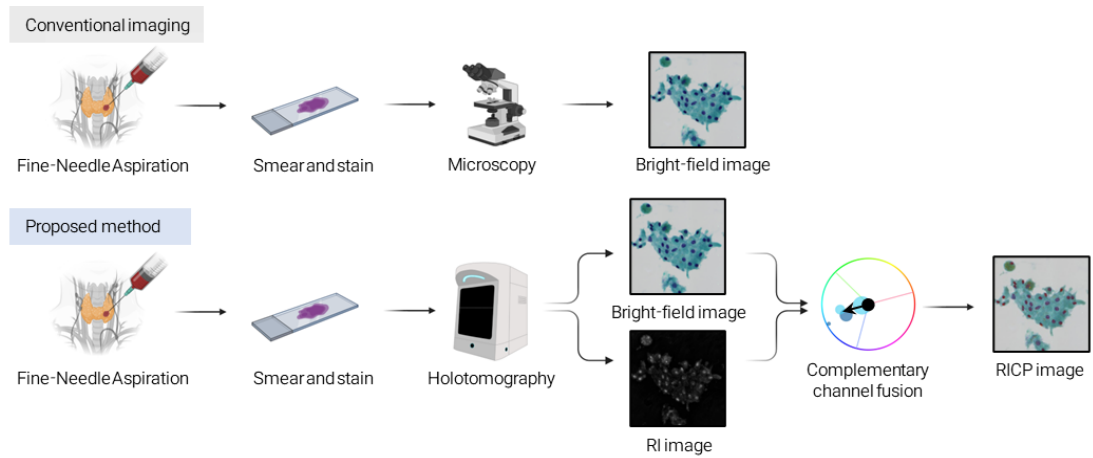

**Figure 1. Comparison between the conventional cytological imaging workflow and the proposed method.** In conventional imaging (top), FNA specimens are smeared, stained, and observed using a BF microscope, yielding color images with limited contrast and potential staining artifacts. In the proposed method (bottom), the same FNAB specimens are imaged by HT, providing both BF and RI maps. The RI information is then integrated into a pseudocoloring process to generate complementary images.

## 2. Methods

*2.1. Thyroid sample preparation*

A single-center cross-sectional study was conducted using thyroid cell clusters obtained via fine-needle aspiration biopsy (FNAB) from patients with benign or malignant thyroid nodules. FNAB slides prepared between July 1 and December 31, 2020, were retrospectively selected from the pathology database of the National Cancer Center (Goyang, Republic of Korea).

Cytological diagnoses were classified according to The Bethesda System for Reporting Thyroid Cytopathology (TBSRTC). Samples were categorized as benign if the FNAB result was reported as "benign (TBSRTC II)". Specimens were classified as malignant if the FNAB result was "malignant (TBSRTC VI)" or "suspicious for malignancy (TBSRTC V)" with subsequent histopathological confirmation of papillary thyroid carcinoma in surgical specimens.

For each patient, a single Papanicolaou (PAP)–stained liquid-based cytology smear slide was selected. An experienced pathologist reviewed all slides and randomly selected up to 20 representative thyroid cell clusters per slide for imaging. Clusters were excluded if they (i) originated from malignant cases but lacked cytologically malignant features or (ii) exhibited insufficient image quality for analysis.

*2.2. Correlative holotomography*

Three-dimensional refractive index (RI) tomograms of thyroid cytology samples were acquired using a HT system based on the digital-micromirror-device (DMD)–illumination holotomography platform (32,33). The system employs a Mach–Zehnder interferometric configuration in which a coherent laser beam (a diode-pumped-solid-state laser, wavelength = 532 nm) is divided by a 2 × 2 single-mode fiber coupler into reference and sample arms. In the sample arm, a series of hologram patterns are projected onto a DMD to control the illumination angle precisely and rapidly (34). The first-order diffracted beam is selected through a spatial filter and relayed onto the specimen via a condenser lens and a high-numerical-aperture objective (UPLSAPO 60× W, NA = 1.2, Olympus). The transmitted field is collected by another objective and interfered with the reference beam on a CMOS detector to record spatially modulated holograms.

For each tomogram, 101 holograms corresponding to multiple illumination angles (100 oblique and one normal incidence, up to 49° in the medium) were recorded. The complex optical fields were retrieved from the holograms, and the 3-D RI distribution was reconstructed using the diffraction-tomography algorithm under the Rytov approximation (35,36). Missing-cone information due to the finite numerical apertures was compensated using a non-negativity iterative constraint algorithm (37).

To enable simultaneous acquisition of both quantitative RI tomograms and color bright-field (BF) images, an RGB LED illumination module was incorporated coaxially into the optical path. The RGB LED array provides spectrally balanced illumination for true-color imaging of Papanicolaou-stained cytology smears, allowing direct correlation between the BF color image and the reconstructed RI tomogram within the same field of view.

## 3. Results

### 3.1. Correlative holotomography and bright-field imaging of thyroid cytology

To obtain both the BF and HT images from a cytological slide, we employed a customized correlative HT system integrated with a BF imaging module (Fig. 2a) (38). The system enables the simultaneous acquisition of quantitative 3D RI tomograms and conventional BF images from the same field of view. FNA smears from thyroid nodules were imaged without additional sample preparation beyond standard staining and mounting, ensuring direct compatibility with routine cytological workflows (see Methods).

Representative BF images (Fig. 2b) clearly depict overall cell morphology, including cytoplasmic contours and nuclear boundaries, yet their diagnostic interpretation is often hindered by staining variability and limited contrast. In contrast, the corresponding RI tomograms (Fig. 2c) obtained by HT provide label-free, quantitative information on intracellular RI distributions, revealing detailed subcellular architectures such as nucleoli and chromatin-rich regions. The measured RI values ranged from 1.337 to 1.363, consistent with the expected range for cytoplasmic and nuclear components in epithelial cells (39–42).

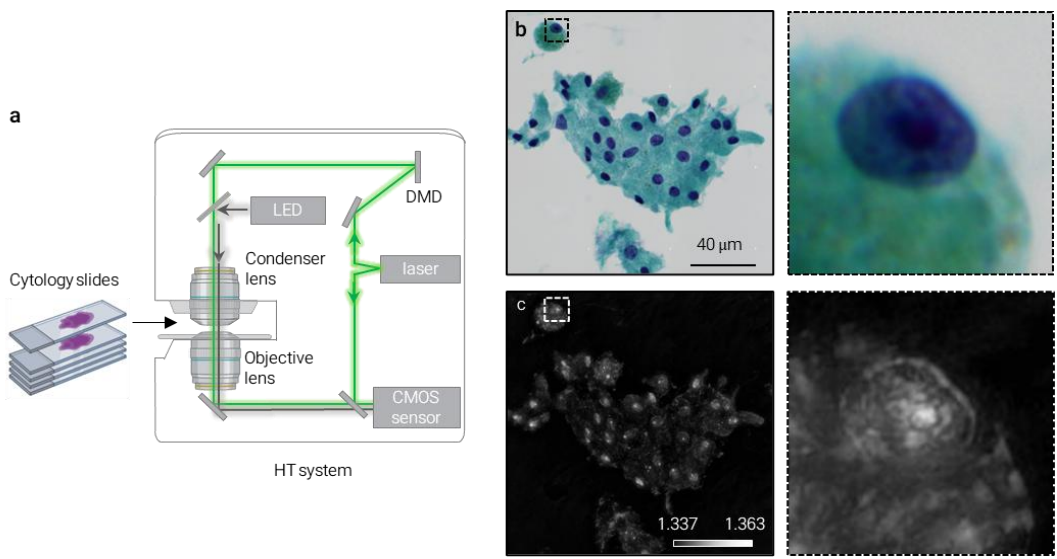

**Figure 2. Optical setup and representative images of thyroid cytology using HT. a,** Schematic of the correlative HT system. Two types of images (BF and RI images) of thyroid cell clusters were taken from cytology slides using the coherence HT system equipped with a brightfield microscope. **b,** Representative BF images of a thyroid fine-needle aspiration sample, showing cytoplasmic and nuclear morphology. **c,** Corresponding RI tomogram from the same region reveals detailed subcellular architecture, including nuclear boundaries and chromatin-rich regions, with quantitative RI values ranging from 1.337 to 1.363.

*3.2. Algorithmic framework for refractive index–correlated pseudocoloring (RICP)*

To integrate quantitative RI information into conventional BF cytology images, we developed a RICP algorithm that enhances subcellular contrast while maintaining the overall color tone of stained specimens (Fig. 3). The RICP method fuses a color BF image and a grayscale RI tomogram into a single pseudocolor image that preserves the BF appearance while embedding RI-derived structural contrast.

The algorithm proceeds in four computational steps.

**Step 1:** The BF image $I_{BF}$ is transformed from RGB to HSV space to decouple chromaticity from intensity. Each pixel is represented by its hue angle $\theta(x, y)$ and saturation magnitude, allowing the color distribution to be expressed on the hue circle. To bias the color mapping toward diagnostically relevant regions, the pixel-wise RI values $I_{RI}$ are used as weights in computing the circular mean hue:

$$\langle c \rangle = \arg\left(\frac{\sum_{x,y} I_{RI}(x, y) e^{i\theta(x,y)}}{\sum_{x,y} I_{RI}(x, y)}\right),$$

which shifts the mean hue toward high-RI regions such as nuclei and nucleoli while suppressing low-RI background pixels.

**Step 2:** A complementary color basis $\{C, O_1, O_2\}$. is constructed, where the complementary hue: $C = \langle c \rangle + \pi \pmod{2\pi}$ represents the least-occupied hue region in the BF image, and $O_1$ and $O_2$ are orthogonal hues spaced $2\pi/3$ apart. The BF image is then projected onto this basis to produce three grayscale channel images $\{I_C, I_{O_1}, I_{O_2}\}$. Because $C$ is minimally represented in the BF color distribution, $I_C$ serves as an optimal channel for incorporating RI-derived information with minimal interference from existing color cues.

**Step 3:** The RI image is normalized to [0,1] and optionally gamma-corrected ($I_{RI} \rightarrow I_{RI}^{\gamma}$) to emphasize contrast in desired RI ranges. A pixelwise weighted combination is then performed to generate the RI-augmented component:

$$I_{C_{RI}} = I_C + (1 - I_C) I_{RI}^{\gamma},$$

which selectively enhances high-RI structures while preserving background brightness.

**Step 4:** Finally, the three channel images $\{I_{C_{RI}}, I_{O_1}, I_{O_2}\}$ are recombined to yield the RICP image. This process retains the appearance of the original BF image while embedding RI-based subcellular contrast within the complementary hue channel.

Representative thyroid cytology images (Fig. 3b) demonstrate that RICP enhances nuclear and subnuclear visibility, revealing chromatin and nucleolar morphology that are indistinct in conventional BF microscopy. Collectively, the RICP framework provides a robust computational approach for fusing quantitative RI data with traditional color cytology, thereby enabling diagnostically meaningful visualization of subtle intracellular heterogeneity.

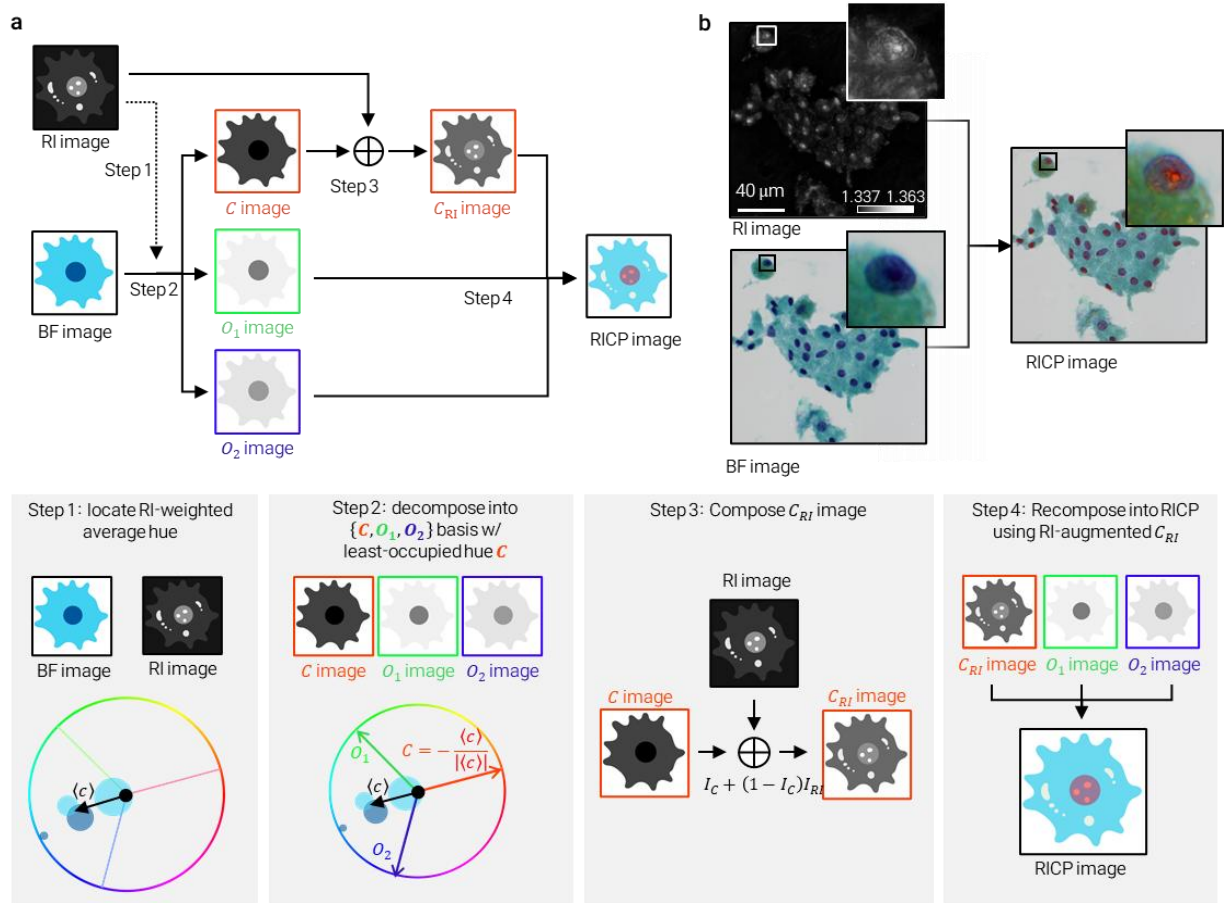

**Figure 3. Conceptual workflow of RICP. a,** (top) Overview of the RICP generation process combining the BF image and RI tomogram to produce the final RICP image. (bottom) Stepwise illustration of the algorithm. *Step 1:* Compute the RI-weighted average hue $\langle c \rangle$ from the BF and RI images. *Step 2:* Decompose the BF image into a color basis set $\{C, O_1, O_2\}$, where $C$ denotes the least-occupied hue. *Step 3:* Generate the RI-weighted color component $C_{RI}$ by combining the intensity of the $C$ channel with the RI map. *Step 4:* Recompose the image into the final RICP using the RI-augmented CICP, enhancing contrast in nuclear regions while maintaining the cytoplasmic color tone. **b,** Representative thyroid cytology images demonstrate how the RICP process reveals chromatin and subnuclear details not visible in the original BF image.

*3.3. Visualization of subcellular morphology in thyroid cytology using RICP*

The RICP method enables enhanced visualization of subcellular structures in stained thyroid cytology specimens compared to conventional BF microscopy (Fig. 4). By integrating quantitative refractive index (RI) information with BF color tone, RICP produces composite images that highlight diagnostically relevant intracellular features while preserving the familiar cytological appearance. This fusion bridges conventional staining with label-free optical contrast, thereby linking morphological cues to intrinsic biophysical properties.

In benign follicular epithelial clusters (Fig. 4a), RICP distinctly reveals intracellular LDs (yellow arrows) and chromatin-dense nucleoli (red arrows) that appear faint or unresolved in BF images. These regions exhibit locally elevated RI values, consistent with the increased molecular density of lipid and nucleoprotein structures. The axial z-stack reconstruction (Fig. 4b) demonstrates the continuity of these organelles across depth ($\Delta z \approx 0.2$ μm), confirming the volumetric imaging capability of HT-based RICP and its ability to capture fine subcellular topology in three dimensions.

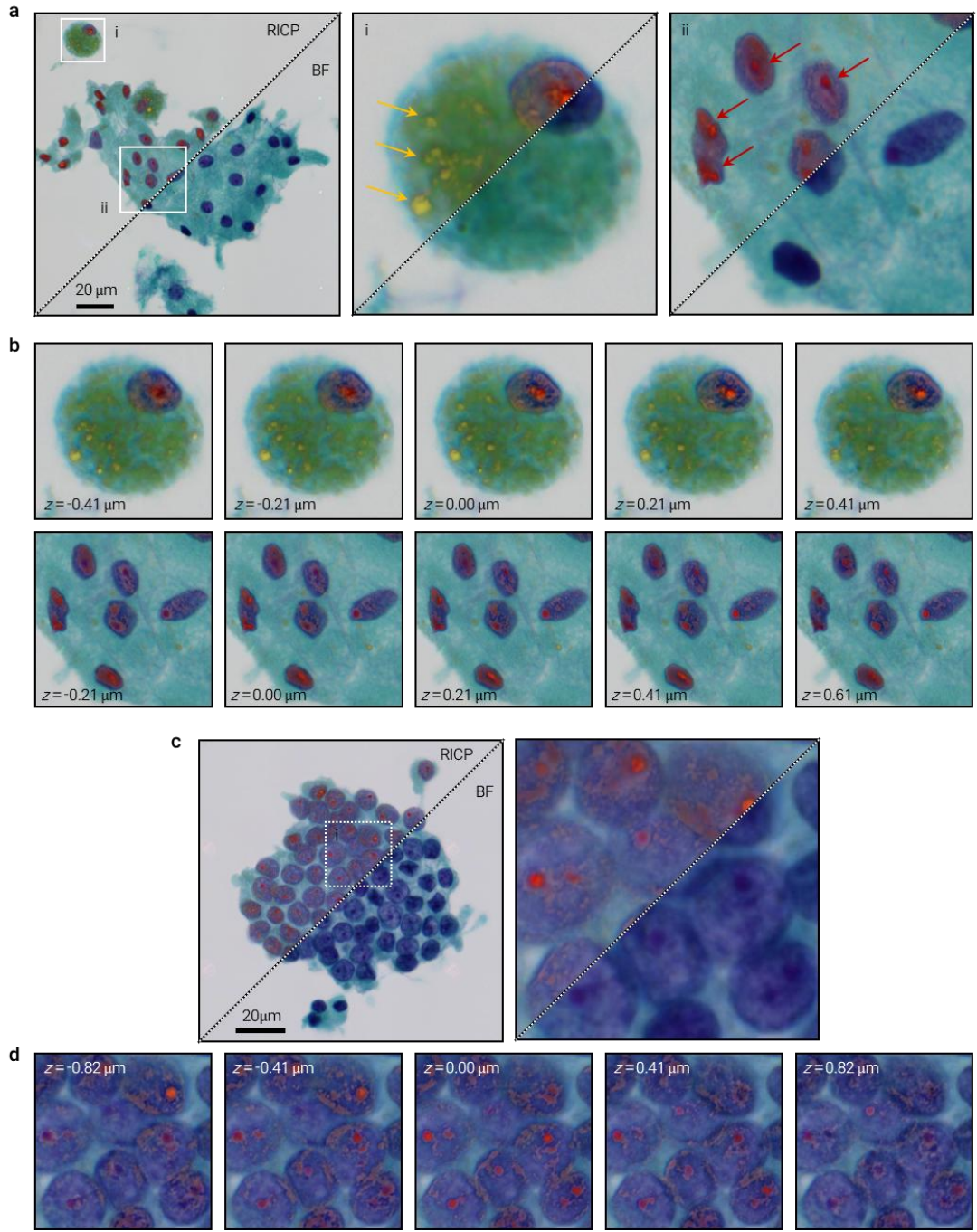

**Figure 4. Visualization of subcellular structures in thyroid cytology using RICP. a,** Side-by-side comparison of BF and RICP images showing two representative regions (i, ii) within a benign follicular epithelial cluster. The RICP image distinctly highlights intracellular lipid droplets (yellow arrows) and chromatin-dense nucleoli (red arrows) that are less discernible in the BF image. **b,** z-slice series of the area (i) reconstructed RICP images from tomographic data demonstrate the 3D continuity of the subcellular features across depth (Δz ≈ 0.2 μm). **c,** Additional examples from a malignant thyroid sample exhibits irregular and thickened nuclear membranes, smaller and more numerous nucleoli with higher RI, and a reduced cytoplasm-to-nucleus ratio. Magnified region on the right corresponds to the boxed area. **d,** Axial RICP slices of the malignant region from (c), revealing the spatial distribution of nucleoli and heterogeneous nuclear architecture across depth, demonstrating RICP's potential for three-dimensional cytological assessment.

By contrast, malignant thyroid clusters display markedly altered subcellular morphology (Fig. 4c). The RICP images reveal thickened and irregular nuclear membranes, smaller yet more numerous nucleoli with higher RI values, and a reduced cytoplasm-to-nucleus ratio—hallmarks of cytological malignancy. Axial reconstructions of the same region (Fig. 4d) visualize the spatial distribution of nucleoli and heterogeneous nuclear texture across depth, providing insight into the 3D organization of chromatin and nucleolar components.

These results demonstrate that RICP complements conventional cytology by coupling morphological and quantitative optical information. This approach enhances the visualization of diagnostically critical features—such as nucleolar morphology, lipid droplet localization, and nuclear irregularity—within a volumetric and label-free framework, offering a new dimension for high-fidelity cytopathological evaluation.

*3.4. Quantitative hue-space analysis distinguishing benign and malignant thyroid cytology*

To evaluate the quantitative capability of RICP in distinguishing cytological phenotypes, we analyzed the color-channel and hue-space distributions of benign and malignant thyroid FNA specimens (Fig. 5). In the RICP color basis $\{I_{C_{RI}}, I_{O_1}, I_{O_2}\}$, the $C$ channel represents the least-occupied hue within the original BF color distribution, validating its suitability for RI fusion (Fig. 5a). After fusion, the $C_{RI}$ component exhibits selectively increased intensity at nuclear membranes and nucleoli—regions known to exhibit high refractive index values—while $O_1$ and $O_2$ primarily retain cytoplasmic and background color information. This channel separation allows the RI-derived contrast to be embedded without distorting the natural BF appearance.

To quantify the global color shift introduced by RICP, we projected each pixel's hue–saturation coordinates into a 2D density map (Fig. 5b). In benign clusters, the hue distributions form narrow, well-defined clusters centered near the dominant cytoplasmic color, reflecting morphological uniformity and consistent RI profiles across nuclei. In contrast, malignant clusters exhibit broadened, asymmetric hue distributions with a directional shift toward the complementary hue (indicated by orange arrows), corresponding to the RI-enhanced nuclear regions. This pattern reflects the higher structural heterogeneity of malignant cells, including irregular chromatin organization and variable nucleolar density.

These results establish that RICP not only enhances qualitative visualization but also provides a quantitative framework for differentiating cytological categories. The hue-space distribution functions as a multidimensional descriptor that correlates color and refractive index heterogeneity—offering a potential basis for automated cytological classification.

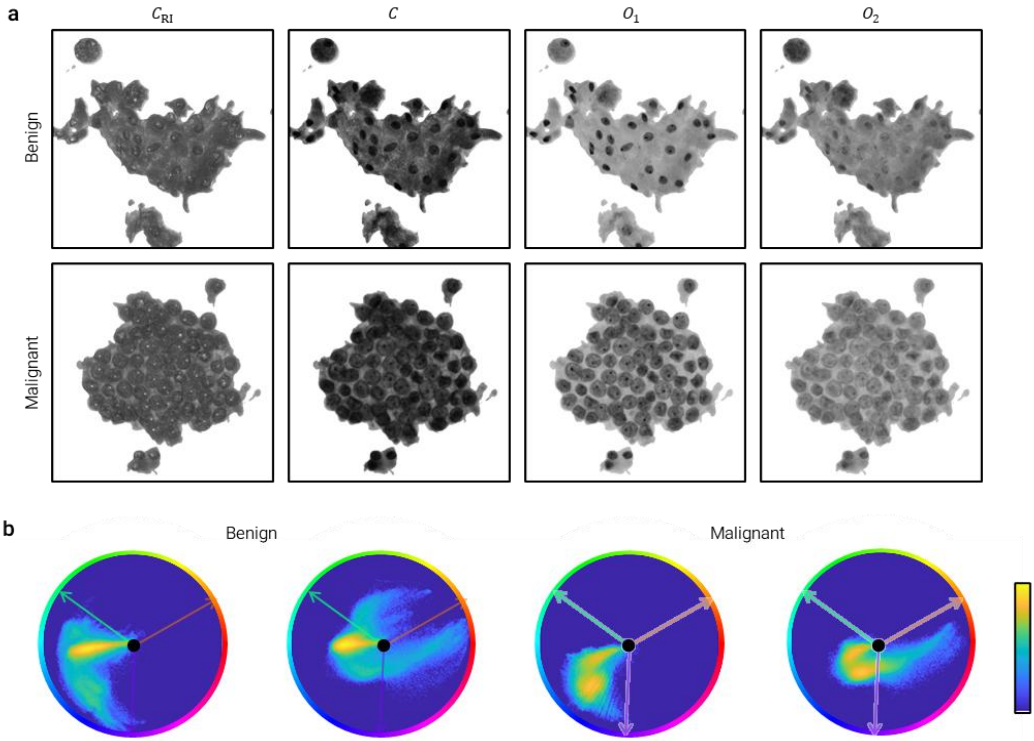

**Figure 5. Quantitative hue-space characterization of benign and malignant thyroid cytology using RICP. (a)** Grayscale channel images corresponding to $C_{RI}$, $C$, $O_1$, and $O_2$ for representative benign (top) and malignant (bottom) thyroid FNA clusters. **(b)** Hue–saturation density plots for BF and RICP images of benign and malignant samples. The RICP images display broadened hue distributions with a directional shift toward the complementary hue (dotted orange arrow), indicating successful embedding of RI-derived contrast into the color space.

## Discussion and Conclusion

In this study, we introduced the RICP framework that integrates quantitative HT data with conventional BF cytology images to enhance subcellular visualization and diagnostic interpretability. By embedding RI–derived contrast into the complementary color channel, RICP provides a unified color representation that preserves the natural appearance of stained specimens while revealing biophysical variations invisible to standard microscopy.

Traditional cytology relies primarily on colorimetric staining and visual pattern recognition to assess nuclear morphology and chromatin distribution. However, these assessments are limited by staining variability and subjective interpretation. HT, on the other hand, provides quantitative contrast based on intrinsic RI distributions that reflect local biomolecular density and composition (43,44). By fusing these complementary modalities, RICP bridges physical and morphological domains—enabling a richer description of cellular architecture that links optical properties with histopathological features.

Application of RICP to thyroid FNA samples demonstrated substantial improvements in visualizing diagnostically critical structures. In benign follicular epithelial clusters, RICP enhanced the visibility of LDs and chromatin-dense nucleoli, confirming their distinct optical signatures. In malignant samples, the method revealed hallmark features of cytological malignancy—irregular and thickened nuclear membranes, increased nucleolar density, and reduced cytoplasm-to-nucleus ratio—with 3D continuity across optical sections. These findings indicate that RICP can augment standard cytopathological interpretation by highlighting subtle intracellular heterogeneity and structural irregularity that often accompany neoplastic transformation.

Beyond qualitative enhancement, RICP provides a quantitative representation of cellular heterogeneity through hue–saturation analysis. The hue-space distributions derived from RICP images distinguish benign from malignant phenotypes: benign cells exhibit compact, well-clustered hue distributions, while malignant cells display broadened, asymmetric hue dispersion aligned with the complementary color axis. This shift arises from RI-enhanced nuclear regions and reflects underlying variations in chromatin organization. Such quantitative descriptors could serve as the basis for automated cytological classification or machine-learning–based diagnostic models, integrating colorimetric and physical contrast into a unified analytical space.

The RICP concept is broadly applicable to various cytological and histological preparations beyond thyroid specimens. Because it operates on digitally acquired BF and RI images, the method is compatible with existing imaging workflows and can be implemented as a post-acquisition computational enhancement. Future work will focus on large-scale validation using diverse pathological samples and integration with AI-driven feature extraction to establish robust, quantitative diagnostic metrics (45). Additionally, the framework may extend to fluorescence–BF or Raman–BF fusion, providing a generalized computational approach for multimodal microscopy (46–50).

While RICP successfully enhances visual and quantitative interpretability, its performance depends on the accuracy of RI reconstruction and the consistency of color calibration in BF imaging. Variations in staining, illumination, or sample thickness could influence the hue mapping process (51–55). Further optimization of normalization and color-space transformation parameters will improve reproducibility across laboratories. Nonetheless, the results presented here demonstrate that RI-correlated pseudocoloring provides a powerful new dimension for cytological imaging—uniting the precision of quantitative optics with the familiarity of conventional morphology.

In conclusion, we developed and validated the RICP framework that integrates HT-derived RI maps with BF cytology for enhanced visualization and quantitative assessment of thyroid FNAB specimens. By embedding RI-based structural contrast into the color domain, RICP bridges physical and morphological information, revealing diagnostically relevant subcellular features such as lipid droplets, nucleoli, and nuclear irregularities. The hue-space analysis further establishes a quantitative metric that differentiates benign and malignant cells based on color–RI correlation. This perceptually grounded and label-free imaging framework introduces a broadly applicable computational strategy for multimodal cytology and lays the foundation for future AI-driven diagnostic systems that combine quantitative optics with traditional morphological evaluation.

## Author Contributions

M.L. designed the study. Data acquisition was done by Y.K.L., S.Y.P., H.L. and E.K.L. M. L., G. K., T. L., and J. P. performed data processing and analysis. All the authors wrote the manuscript.

## Ethics Statement

This study was conducted according to the guidelines of the Declaration of Helsinki and was approved by the institutional review board of the National Cancer Center (IRB number: NCC2020-0126), which waived the requirement for informed consent for this study.

## Conflicts of Interest

G.K. and Y.KP. have financial interests in Tomocube Inc., a company that commercializes HT system. The remaining authors declare no competing interests.

## Data Availability Statement

The data that support the findings of this study are available upon reasonable request from the authors.

**References**

1. Ravetto C, Colombo L, Dottorini ME. Usefulness of fine-needle aspiration in the diagnosis of thyroid carcinoma. Cancer Cytopathology 2000;90:357–363.

2. Mertz J. Introduction to Optical Microscopy. Cambridge University Press; 2019. 475 p.

3. Park Y, Depeursinge C, Popescu G. Quantitative phase imaging in biomedicine. Nature Photon 2018;12:578–589.

4. Kim G, Hugonnet H, Kim K, Lee J-H, Lee SS, Ha J, Lee C, Park H, Yoon K-J, Shin Y. Holotomography. Nature Reviews Methods Primers 2024;4:51.

5. Pirone D, Lim J, Merola F, Miccio L, Mugnano M, Bianco V, Cimmino F, Visconte F, Montella A, Capasso M, Iolascon A, Memmolo P, Psaltis D, Ferraro P. Stain-free identification of cell nuclei using tomographic phase microscopy in flow cytometry. Nat. Photon. 2022;16:851–859.

6. Kim H, Oh S, Lee S, Lee K suk, Park Y. Recent advances in label-free imaging and quantification techniques for the study of lipid droplets in cells. Current Opinion in Cell Biology 2024;87:102342.

7. Khadem H, Mangini M, Ferrara MA, De Luca AC, Coppola G. Polarization-Sensitive Holotomography for Multidimensional Label-Free Imaging and Characterization of Lipid Droplets in Cancer Cells. Advanced Science n/a:e09420.

8. Zdańkowski P, Winnik J, Rogalski M, Marzejon MJ, Wdowiak E, Dudka W, Józwik M, Trusiak M. Polarization Gratings Aided Common-Path Hilbert Holotomography for Label-Free Lipid Droplets Content Assay. Laser & Photonics Reviews 2025;19:2401474.

9. Kim T, Yoo J, Do S, Hwang DS, Park Y, Shin Y. RNA-mediated demixing transition of low-density condensates. Nat Commun 2023;14:2425.

10. Esposito M, Fang C, Cook KC, Park N, Wei Y, Spadazzi C, Bracha D, Gunaratna RT, Laevsky G, DeCoste CJ, Slabodkin H, Brangwynne CP, Cristea IM, Kang Y. TGF-β-induced DACT1 biomolecular condensates repress Wnt signalling to promote bone metastasis. Nat Cell Biol 2021;23:257–267.

11. Kawagoe S, Matsusaki M, Mabuchi T, Ogasawara Y, Watanabe K, Ishimori K, Saio T. Mechanistic Insights Into Oxidative Response of Heat Shock Factor 1 Condensates. JACS Au 2025;5:606–617.

12. Lee SY, Park HJ, Best-Popescu C, Jang S, Park YK. The Effects of Ethanol on the Morphological and Biochemical Properties of Individual Human Red Blood Cells. PLOS ONE 2015;10:e0145327.

13. Anon. Deep-learning-based three-dimensional label-free tracking and analysis of immunological synapses of CAR-T cells | eLife. Available at: https://elifesciences.org/articles/49023. Accessed October 9, 2025.

14. Sung M, Kim JH, Min H-S, Jang S, Hong J, Choi BK, Shin J, Chung KS, Park YR. Three-dimensional label-free morphology of CD8 + T cells as a sepsis biomarker. Light Sci Appl 2023;12:265.

15. Lee MJ, Kim G, Lee MS, Shin JW, Lee JH, Ryu DH, Kim YS, Chung Y, Kim K, Park Y. Functional immune state classification of unlabeled live human monocytes using holotomography and machine learning. 2025:2023.09.12.557503. Available at: https://www.biorxiv.org/content/10.1101/2023.09.12.557503v2. Accessed October 9, 2025.

16. Kim H, Kim G, Park H, Lee MJ, Park Y, Jang S. Integrating holotomography and deep learning for rapid detection of NPM1 mutations in AML. Sci Rep 2024;14:23780.

17. Hugonnet H, Kim YW, Lee M, Shin S, Hruban RH, Hong S-M, Park Y. Multiscale label-free volumetric holographic histopathology of thick-tissue slides with subcellular resolution. AP 2021;3:026004.

18. Liu Y, Uttam S. Perspective on quantitative phase imaging to improve precision cancer medicine. JBO 2024;29:S22705.

19. Lee MJ, Lee J, Ha J, Kim G, Kim H-J, Lee S, Koo B-K, Park Y. Long-term three-dimensional high-resolution imaging of live unlabeled small intestinal organoids via low-coherence holotomography. Exp Mol Med 2024;56:2162–2170.

20. Park D, Lee D, Kim Y, Park Y, Lee Y-J, Lee JE, Yeo M-K, Kang M-W, Chong Y, Han SJ, Choi J, Park J-E, Koh Y, Lee J, Park Y, Kim R, Lee JS, Choi J, Lee S-H, Ku B, Kang DH, Chung C. Cryobiopsy: A Breakthrough Strategy for Clinical Utilization of Lung Cancer Organoids. Cells 2023;12:1854.

21. Li S, Kang X, Fang L, Hu J, Yin H. Pixel-level image fusion: A survey of the state of the art. Information Fusion 2017;33:100–112.

22. James AP, Dasarathy BV. Medical image fusion: A survey of the state of the art. Information Fusion 2014;19:4–19.

23. Liu Y, Chen X, Ward RK, Jane Wang Z. Image Fusion With Convolutional Sparse Representation. IEEE Signal Processing Letters 2016;23:1882–1886.

24. Peixoto HM, Munguba H, Cruz RM, Guerreiro AM, Leao RN. Automatic tracking of cells for video microscopy in patch clamp experiments. BioMedical Engineering OnLine 2014;13:78.

25. Glatz J, Symvoulidis P, Garcia-Allende PB, Ntziachristos V. Robust overlay schemes for the fusion of fluorescence and color channels in biological imaging. JBO 2014;19:040501.

26. Watson JR, Gainer CF, Martirosyan N, M.d JS, Jr GML, M.d RA, Romanowski M. Augmented microscopy: real-time overlay of bright-field and near-infrared fluorescence images. Available at: https://www.spiedigitallibrary.org/journals/journal-of-biomedical-optics/volume-20/issue-10/106002/Augmented-microscopy--real-time-overlay-of-bright-field-and/10.1117/1.JBO.20.10.106002.full. Accessed October 9, 2025.

27. Tu T-M, Su S-C, Shyu H-C, Huang PS. A new look at IHS-like image fusion methods. Information Fusion 2001;2:177–186.

28. Vivone G, Alparone L, Chanussot J, Dalla Mura M, Garzelli A, Licciardi GA, Restaino R, Wald L. A Critical Comparison Among Pansharpening Algorithms. IEEE Transactions on Geoscience and Remote Sensing 2015;53:2565–2586.

29. Witzel C, Gegenfurtner KR. Color Perception: Objects, Constancy, and Categories. Annual Review of Vision Science 2018;4:475–499.

30. Jin X, Chen G, Hou J, Jiang Q, Zhou D, Yao S. Multimodal sensor medical image fusion based on nonsubsampled shearlet transform and S-PCNNs in HSV space. Signal Processing 2018;153:379–395.

31. Cui C, Yang H, Wang Y, Zhao S, Asad Z, Coburn LA, Wilson KT, Landman BA, Huo Y. Deep multimodal fusion of image and non-image data in disease diagnosis and prognosis: a review. Prog. Biomed. Eng. 2023;5:022001.

32. Anon. Optical diffraction tomography using a digital micromirror device for stable measurements of 4D refractive index tomography of cells. Available at: https://www.spiedigitallibrary.org/conference-proceedings-of-spie/9718/971814/Optical-diffraction-tomography-using-a-digital-micromirror-device-for-stable/10.1117/12.2216769.full. Accessed October 3, 2025.

33. Lee K, Kim K, Kim G, Shin S, Park Y. Time-multiplexed structured illumination using a DMD for optical diffraction tomography. Opt. Lett., OL 2017;42:999–1002.

34. Shin S, Kim K, Yoon J, Park Y. Active illumination using a digital micromirror device for quantitative phase imaging. Opt. Lett. 2015;40:5407.

35. Kak AC, Slaney M, Wang G. Principles of Computerized Tomographic Imaging. Medical Physics 2002;29:107–107.

36. Wolf E. Three-dimensional structure determination of semi-transparent objects from holographic data. Optics Communications 1969;1:153–156.

37. Lim J, Lee K, Jin KH, Shin S, Lee S, Park Y, Ye JC. Comparative study of iterative reconstruction algorithms for missing cone problems in optical diffraction tomography. Opt. Express 2015;23:16933.

38. Lee YK, Ryu D, Kim S, Park J, Park SY, Ryu D, Lee H, Lim S, Min H-S, Park Y, Lee EK. Machine-learning-based diagnosis of thyroid fine-needle aspiration biopsy synergistically by Papanicolaou staining and refractive index distribution. Sci Rep 2023;13:9847.

39. Kim EH, Park S, Kim YK, Moon M, Park J, Lee KJ, Lee S, Kim Y-P. Self-luminescent photodynamic therapy using breast cancer targeted proteins. Science Advances 2020;6:eaba3009.

40. Aldonza MBD, Reyes RDD, Kim YS, Ku J, Barsallo AM, Hong J-Y, Lee SK, Ryu HS, Park Y, Cho J-Y, Kim Y. Chemotherapy confers a conserved secondary tolerance to EGFR inhibition via AXL-mediated signaling bypass. Sci Rep 2021;11:8016.

41. Park C, Lee K, Baek Y, Park Y. Low-coherence optical diffraction tomography using a ferroelectric liquid crystal spatial light modulator. Opt. Express, OE 2020;28:39649–39659.

42. Lee AJ, Hugonnet H, Park W, Park Y. Three-dimensional label-free imaging and quantification of migrating cells during wound healing. Biomed. Opt. Express, BOE 2020;11:6812–6824.

43. Barer R. Determination of dry mass, thickness, solid and water concentration in living cells. Nature 1953;172:1097–1098.

44. Popescu G, Park Y, Lue N, Best-Popescu C, Deflores L, Dasari RR, Feld MS, Badizadegan K. Optical imaging of cell mass and growth dynamics. American Journal of Physiology-Cell Physiology 2008;295:C538–C544.

45. Park J, Bai B, Ryu D, Liu T, Lee C, Luo Y, Lee MJ, Huang L, Shin J, Zhang Y. Artificial intelligence-enabled quantitative phase imaging methods for life sciences. Nature methods 2023;20:1645–1660.

46. Chowdhury S, Eldridge WJ, Wax A, Izatt JA. Structured illumination microscopy for dual-modality 3D sub-diffraction resolution fluorescence and refractive-index reconstruction. Biomed. Opt. Express, BOE 2017;8:5776–5793.

47. Shin S, Kim D, Kim K, Park Y. Super-resolution three-dimensional fluorescence and optical diffraction tomography of live cells using structured illumination generated by a digital micromirror device. Sci Rep 2018;8:9183.

48. Liu C, Malek M, Poon I, Jiang L, Siddiquee AM, Sheppard CJR, Roberts A, Quiney H, Zhang D, Yuan X, Lin J, Depeursinge C, Marquet P, Kou SS. Simultaneous dual-contrast three-dimensional imaging in live cells via optical diffraction tomography and fluorescence. Photon. Res., PRJ 2019;7:1042–1050.

49. Anon. Protein and lipid mass concentration measurement in tissues by stimulated Raman scattering microscopy | PNAS. Available at: https://www.pnas.org/doi/abs/10.1073/pnas.2117938119. Accessed October 9, 2025.

50. Lee M, Hugonnet H, Park Y. Inverse problem solver for multiple light scattering using modified Born series. Optica, OPTICA 2022;9:177–182.

51. Ban S, Min E, Ahn Y, Popescu G, Jung W. Effect of tissue staining in quantitative phase imaging. Journal of Biophotonics 2018;11:e201700402.

52. Lim J, Wahab A, Park G, Lee K, Park Y, Ye JC. Beyond Born-Rytov limit for super-resolution optical diffraction tomography. Opt. Express, OE 2017;25:30445–30458.

53. Oh C, Hugonnet H, Lee M, Park Y. Digital aberration correction for enhanced thick tissue imaging exploiting aberration matrix and tilt-tilt correlation from the optical memory effect. Nat Commun 2025;16:1685.

54. Hugonnet H, Oh C, Park J, Park Y. Pupil phase series: a fast, accurate, and energy-conserving model for forward and inverse light scattering in thick biological samples. Opt. Express, OE 2025;33:34255–34266.

55. Kim J, Bolton B, Moshksayan K, Khanna R, Swartz ME, Ziemczonok M, Kamra M, Jorn KA, Parekh SH, Kujawińska M, Eberhart J, Cenik ES, Ben-Yakar A, Chowdhury S. Inverse-scattering in biological samples via beam-propagation. 2025:2025.08.17.670744. Available at: https://www.biorxiv.org/content/10.1101/2025.08.17.670744v1. Accessed October 9, 2025.